\documentclass[12pt]{article}

  \usepackage{graphicx}
  \usepackage{epsfig}
  \usepackage{amssymb}
  \usepackage{subfigure}

\evensidemargin - .5truecm
\begin{document}

\newcommand{\be}{\begin{equation}}
\newcommand{\ee}{\end{equation}}
\newcommand{\nn}{\nonumber}
\newcommand{\bea}{\begin{eqnarray}}
\newcommand{\eea}{\end{eqnarray}}
\newcommand{\bfig}{\begin{figure}}
\newcommand{\efig}{\end{figure}}
\newcommand{\bc}{\begin{center}}
\newcommand{\ec}{\end{center}}


\begin{titlepage}
\nopagebreak
{\flushright{
        \begin{minipage}{5cm}
         LPSC 10/95\\
         RM3-TH/10-15 \\
         IFUM-960-FT \\ 
        \end{minipage}        }

}
\renewcommand{\thefootnote}{\fnsymbol{footnote}}
\vskip 2.0cm
\begin{center}
\boldmath
{\Large\bf On the }\\[9pt]
{\Large\bf Generalized Harmonic Polylogarithms }\\[7pt]
{\Large\bf of One Complex Variable}\unboldmath
\vskip 1.cm
{\large  R.~Bonciani\footnote{Email:
Roberto.Bonciani@lpsc.in2p3.fr}}
\vskip .2cm
{\it 
LPSC,
Universit\'e Joseph Fourier/CNRS-IN2P3/INPG,\\
F-38026 Grenoble, France}
\vskip .4cm
{\large G.~Degrassi\footnote{Email: degrassi@fis.uniroma3.it}},
\vskip .2cm
{\it Dipartimento di Fisica, Universit\`a di Roma Tre and 
INFN, Sezione di Roma III, \\ 
I-00146 Rome, Italy} 
\vskip .4cm
{\large A.~Vicini\footnote{Email: Alessandro.Vicini@mi.infn.it}}
\vskip .2cm
{\it Universit\`a degli Studi di Milano and
INFN, Sezione di Milano,\\
I-20133 Milano, Italy} 
\end{center}
\vskip 1.5cm

\begin{abstract}
We describe how to compute numerically in the complex plain a set of 
Generalized Harmonic Polylogarithms (GHPLs) with square roots
in the weights, using the {\tt C++/GiNaC} numerical routines of Vollinga 
and Weinzierl. As an example, we provide the numerical values of the NLO 
electroweak light-fermion corrections to the Higgs boson production in 
gluon fusion in the case of complex $W$ and $Z$ masses.
\vskip .4cm
{\it Key words}: Harmonic Polylogarithms, Feynman diagrams, Multi-loop 
calculations, Higgs physics
\end{abstract}
\vfill
\end{titlepage}    
\setcounter{footnote}{0}

\section{Introduction}

In recent years, the need of accurate theoretical predictions for scattering
amplitudes in collider physics requested a strong effort in the development of methods
and strategies for the calculation of multi-loop Feynman diagrams. In particular, it
was recently possible to afford many analytic calculations  unthinkable up to fifteen
years ago. 

This success is largely due to reliable and powerful algorithms, as for instance the
so-called ``Laporta algorithm'' \cite{Laportaalgorithm}.  The calculation of a
physical observable in perturbation theory requires the (numerical or analytical)
evaluation of a  large number of regularized scalar integrals. The Laporta algorithm
allows to  reduce this large number of scalar integrals to a linear combination of a
small  set of independent scalar integrals, called the ``Master Integrals'' (MIs) of 
the problem under consideration. The Laporta algorithm is based on the
Integration-by-Parts Identities (IBPs) \cite{Tka}, a set of relations that link scalar
integrals with a different power of the propagators and of scalar products in the
numerator among each other\footnote{Public implementations of the algorithm are
available in \cite{LAPCODE}.}. The interplay between the ``Differential Equations
Method'' \cite{DiffEq} and the techniques based on  Mellin-Barnes representations of
the integrals \cite{MB} provides, then, a  powerful tool for the analytic calculation
of the MIs.

Another important ingredient for the analytic calculation of the higher-order
corrections to a physical observable is the identification of the base of  functions
in terms of which the MIs can be expressed.
This base of functions is strictly related to the structure of the thresholds of the
Feynman diagrams under consideration.

The connection between Feynman diagrams with a simple structure of thresholds  and the
functional base of the Harmonic Polylogarithms\footnote{Together with  the related
harmonic \cite{HS}, nested \cite{NS} and binomial sums \cite{BS}.} 
(HPLs) \cite{HPLs}  is completely clear.
Feynman diagrams with a richer structure of thresholds require the introduction of
one- \cite{HPLs+,Aglietti:2004tq} and two-dimensional 
\cite{2dHPLs,Birthwright:2004kk,Bonciani:2003cj,Bonciani:2007eh,Bonciani:2008az}   
extensions of the HPLs, that will be generically referred to as  Generalized Harmonic
Polylogarithms (GHPLs).

We briefly recall the advantages of using the (G)HPLs, that constitute a well suited
functional base in which to express the analytic results. 
{\it i)} The structure of the (G)HPLs, is strictly connected with the  solution, with
the Euler method,  of the first-order linear differential equations satisfied by the
Feynman amplitudes.
{\it ii)} The (G)HPLs constitute a base of linearly independent functions. 
{\it iii)} The base of (G)HPLs provides a perfect control on the analytical properties 
of the MIs and, therefore, of the physical observable that we aim to calculate.
{\it iv)} Finally, there are available numerical routines, that allow a precise 
evaluation of the (G)HPLs in FORTRAN \cite{NUMHPL}, Mathematica  \cite{Daniel1}, {\tt
C++} \cite{Weinzierl}. 

In this paper, we provide a detailed analysis of the GHPLs of a single variable, with
weights containing square roots. These GHPLs were introduced in 
\cite{Aglietti:2004tq} for the analytic expression of the MIs concerning the
electroweak form factor \cite{Aglietti:2003yc,Aglietti:2004tq}.  In
\cite{Aglietti:2004nj}, ad hoc numerical routines  were made for the evaluation of a
subset of functions of this class occurring in the calculation of the electroweak NLO
corrections to the production of a Higgs boson in gluon fusion and its decay in two
photons. 
The purpose of the present analysis is  to give a general framework for the evaluation
of  GHPLs with weights containing square roots and to show that the evaluation of all
of the GHPLs introduced in \cite{Aglietti:2004tq} can be performed using already
existing numerical routines.

In section \ref{Def}, we recall the definition and the basic properties of 
the 1-dimensional HPLs and GHPLs with square roots in the weights. We start 
on the case of real variable $x$ and on the subset of GHPLs that have the
following possible weights: $\{ -r,-4,-1,0 \}$. Subsequently we discuss their
analytic continuation.
In section \ref{lin}, we illustrate how to move from the set of GHPLs with
square roots in the weights to a set of generalized polylogarithms, with linear
weights. This is the remark that allows a numerical evaluation of the
GHPLs with square roots, using the {\tt C++} routines of Vollinga and Weinzierl 
\cite{Weinzierl} (in the following they will be referred to as ``VW routines'').
In section \ref{complex}, we consider the case in which the variable $x$ is 
complex. We provide a demonstration of the GHPLs scale 
invariance in the complex plain, justifying the use of the VW
routines also in this case.
In section \ref{general}, we introduce additional weights and discuss the 
transformations of section \ref{lin} applied to this new extended set.
In section \ref{higgs}, we apply the results of this paper to the numerical 
evaluation of the GHPLs involved in the NLO light-fermion electroweak corrections 
to the Higgs boson production in gluon fusion, in the case of complex $W$ and $Z$
masses.
Finally, in the appendices, we provide the analytic expressions of the linearized 
GHPLs involved in the calculation shown in section \ref{higgs}.

\section{HPLs and GHPLs of a Real Variable \label{Def}}

In this section, we recall the definition and the properties of the one-dimensional
Harmonic Polylogarithms (HPLs) of a real variable and their generalization (GHPLs), 
with square roots and linear weights, introduced in \cite{Aglietti:2004tq}.

\subsection{Harmonic Polylogarithms \label{HPLs}}

The set of functions denominated Harmonic Polylogarithms (HPLs) \cite{HPLs} is defined
as repeated integrations of the following three fundamental\footnote{Note a minus sign
in the weight $+1$ with respect to the Remiddi-Vermaseren definition of \cite{HPLs}} 
``weight functions'':
\be
f(-1;t) = \frac{1}{t+1} \, , \qquad
f(0;t) = \frac{1}{t} \, , \qquad
f(1;t) = \frac{1}{t-1} \, .
\label{basicHPLs}
\ee
Note that the functions in Eq.~(\ref{basicHPLs}) have a non-integrable singularity in 
$t=-1$, $t=0$, and $t=1$ respectively. 
The related HPLs of weight 1 are
\bea
H(-1;x) & = & \int_0^x \frac{dt}{t+1} \, = \log{(x+1)} \, , 
\label{w1HPLsm1} \\
H(0;x)  & = & \int_1^x \frac{dt}{t} \,   = \log{(x)}     \, , 
\label{w1HPLs0} \\
H(1;x)  & = & \int_0^x \frac{dt}{t-1} \, = \log{(1-x)} \, ,
\label{w1HPLsp1}
\eea
where $x$ is a real variable ($x \in \mathbb{R}$). Since the logarithms have branch
cuts on the real axis for $x \leq -1$, $x \leq 0$, and $x \geq 1$, respectively, the
three HPLs in  Eqs.~(\ref{w1HPLsm1},\ref{w1HPLs0},\ref{w1HPLsp1}) are real and
uniquely defined only for $x > -1$, $x > 0$, and $x < 1$, respectively. Outside these 
intervals, the logarithms become complex, and a prescription for the approach to the 
branch cut has to be chosen (see section \ref{analytHPLs}).

An HPL with weight 2 or bigger is defined through a repeated integration of the weight
functions of Eq.~(\ref{basicHPLs}). If ${\mathbf w}$ is a vector with $w$ components
consisting of a sequence of $-1$, $0$, and $+1$, we define the HPL of weight $w+1$ as
follows:
\be
H(a,{\mathbf w};x) = \int_0^x dt \, f(a;t) \, H({\mathbf w};t) \, ,
\quad a=-1,0,1 \, ,
\ee
with the exception of the case in which the weights are only zeroes, defined as
\be
H({\mathbf 0}_{w+1};x) = \int_1^x dt \, f(0;t) \, H({\mathbf 0}_w;t) = 
\frac{1}{(w+1) !} \log^{w+1}{x} \, .
\ee

The singularity structure and analyticity properties of the HPLs derives from the
properties of the logarithms. A logarithmic singularity in 0, $-1$ or $+1$ (and
$+\infty$)  can occur, together with the respective branch cuts in $x \geq 1$, $x \leq
0$, and  $x \leq -1$, as discussed in the next section.

The HPLs satisfy a shuffle algebra according to which a product of two HPLs of 
weights $n_1$ and $n_2$ is a combination of HPLs of weight $n = n_1+n_2$. Let 
${\mathbf w}_1$ be a vector with $n_1$ components and ${\mathbf w}_2$ a vector 
with $n_2$ components, both consisting of a sequence of -1, 0 and +1. Then we have:
\be
H({\mathbf w}_1;x) \, H({\mathbf w}_2;x) = \sum_{{\mathbf w}={\mathbf w}_1  
\uplus  {\mathbf w}_2} \! \! H({\mathbf w};x) \, ,
\label{shuffle}
\ee
where ${\mathbf w}$ is a vector with $n = n_1+n_2$ components $-1$, $0$, or $+1$. The 
sign $\uplus$ means that the order in the components of ${\mathbf w}_1$ and ${\mathbf
w}_2$ has to be preserved in the sequence ${\mathbf w}$. 
For instance, an HPL with weight 4 can be made out of a product of two HPLs, one  of
weight 1 and another of weight 3, or both of weight 2. The relevant formulas are:
\bea
\! \! H(a;x) \, H(b,c,d;x) \! \! & = & \! \! H(a,b,c,d;x) \! + \! H(b,a,c,d;x) \! 
+ \! H(b,c,a,d;x) \! + \! H(b,c,d,a;x) \, , \\
\! \! H(a,b;x) \, H(c,d;x) \! \! & = & \! \!   H(a,b,c,d;x) \! + \! H(a,c,b,d;x) \! 
+ H(a,c,d,b;x) \! + \! H(c,a,b,d;x) \nn\\
\! \! \! \! & & \! \! + H(c,a,d,b;x) \! +\!  H(c,d,a,b;x) \, .
\eea
The demonstration that HPLs satisfy the shuffle algebra in Eq.~(\ref{shuffle}) 
can be done by induction, using integration by parts (see \cite{HPLs}).

\subsubsection{Analytic Continuation \label{analytHPLs}}

The HPLs are, in general, complex, depending on the value of the real variable $x$. 
In many relevant physical cases, the calculation of the Feynman integrals involved in
some observable is done in a restricted range of $x$. For instance, if $x$ is related
to the squared center of mass energy $s$ through the relation $x=-s/m^2$, with $m$ a
mass scale of the problem, the Feynman integrals are usually solved in the  so-called
Euclidean region: $x \geq 0$\footnote{If, moreover, $x$ is related to $s$ through a
quadratic relation, as the transformation of variable in Eq.~(\ref{redvar}), the
Euclidean region is even more restricted: $0 \leq x < 1$.}.

Let us suppose, then, $x \geq 0$. In the range $0 \leq x \leq 1$ all the HPLs are
real.  For $x > 1$, instead, we have a possible cut, corresponding to the HPLs with a
+1 in the right-most weight, and therefore, ultimately, to the $\log{(1-x)}$, that has
an imaginary part in this region.

In the case of HPLs of weight 1, depending on the prescription adopted, this imaginary
part is $\pm i \pi$:
\be
H(1;x) = \log{(1-x)} = \log{|1-x|} \pm i \pi \, \theta{(x-1)} \, ,
\label{imxg1}
\ee
while $H(0;x)$ and $H(-1;x)$ are real for positive $x$.

In the case of HPLs of weight 2, or bigger, an explicit expression for the imaginary 
part can be found, using the shuffle algebra properties to move the weights 1 from 
the right to the left in the sequence and the relation of Eq.~(\ref{imxg1}). For 
instance, for $H(0,1;x)=\mbox{Li}_2(x)$, the Euler Dilogarithm, we have:
\bea
H(0,1;x) &=& H(0;x) H(1;x) - H(1,0;x) \, , \nn\\
&=& \log{(x)} \log{|1-x|} - H(1,0;x) 
\pm i \pi \, \theta{(x-1)} \log{(x)} \, .
\label{reldilog}
\eea
The HPL $H(1,0;x)$ is real for $x>1$ and the imaginary part of $H(0,1;x)$ is explicitly 
given by the last term in Eq.~(\ref{reldilog}).

Once the searched analytic expression is known in the range of $x>0$, one has to do 
an analytic transformation to move back to the Minkowski region ($s>0$ and then
$x<0$).  Because of causality, the Mandelstam invariant $s$ has to be assigned a
positive  vanishing imaginary part, $s+i0^{+}$. Therefore, if $x=-s/m^2$, the case
$s>0$ is  recovered using
\be
x \to - x' - i 0^{+} \, ,
\label{ancon}
\ee
where now $x'=s/m^2 > 0$.

If $0 < x' \leq 1$ ($-1 \leq x < 0$), we have to take into account the  branch
cut connected to the weight 0. For HPLs of  weight 1 we have
\bea
H(0;x) &\to & H(0;-x'-i 0^{+}) = H(0;x') - i \pi \, ,
\label{imxg0} \\
H(1;x) & \to & H(1;-x'-i 0^{+}) = H(-1;x') \, , \\
H(-1;x) & \to & H(-1;-x'-i 0^{+}) = H(1;x') \, .
\eea
Therefore, only the $\log{(x)}$ gets the imaginary part.
The case of HPLs of weight 2 or bigger has to be treated extracting, with the help of 
the shuffle algebra, the right-most zeroes (trailing zeroes). The HPLs that have no 
zeroes on the right of the sequence of the weights do not get imaginary parts.
Moreover,  moving from $x$ to $x'$, the weights flip in sign. The $\log^n{x}$
extracted with the  algebra are then transformed according to Eq.~(\ref{imxg0}). As a
simple example,  consider the function $H(1,-1,0,x)$. Using the shuffle algebra we
obtain: 
\be
H(1,-1,0;x) = H(0,-1,1;x) - H(0,-1;x)H(1;x) + H(1,-1;x) H(0;x) \, .
\ee 
The cut behaviour is explicitly extracted as $H(0;x)$. The other functions, 
$H(1;x)$, $H(1,-1;x)$, $H(0,-1;x)$, and $H(0,-1,1;x)$ are real for $-1 \leq x < 0$.
Therefore, using Eqs.~(\ref{ancon}) and (\ref{imxg0}), we have:
\bea
H(1,-1,0;- x' - i 0^{+}) &=&  H(0,-1,1;- x' - i 0^{+}) 
- H(0,-1;- x' - i 0^{+})H(1;- x' - i 0^{+}) \nn\\
& &  
+ H(1,-1;- x' - i 0^{+})H(0;- x' - i 0^{+}) \, , \nn\\
& = & H(0,1,-1;x') - H(0,1;x')H(-1;x') + H(-1,1;x') H(0;x') \nn\\
& & - i \pi \, H(-1,1;x') \, .
\label{ex0to1}
\eea

If $x'>1$ ($x < -1$), we have the superposition of the cuts connected to the
weights  0 and $1$. The first cut is taken into account with the analytic continuation
discussed just above: $x \to - x' - i 0^{+}$ with $0 < x' \leq 1$. Now, if $x'>1$,
the HPLs with the right-most weight 1 exhibits an imaginary part, which comes from
the  $\log{(1-x')}$. Using the shuffle algebra, the weights +1 on the right of the
sequence can be moved to the left, and the $\log{(1-x')}$ explicitly extracted.
Since $s$ is to be understood with a vanishing positive imaginary part, this is also
the case of $x'$: $x'+i0^+$. Therefore, we have:
\be
\log{(1-x'-i0^+)} = \log{(x'-1)} - i \pi \, , \qquad x'> 1 \, .
\ee
Continuing with the example of $H(1,-1,0;x)$, now we start from Eq.~(\ref{ex0to1}).
The cut for $x'>1$ shows up in the functions $H(0,1;x')$ and $H(-1,1;x')$. Using again
the shuffle algebra we can rewrite these functions extracting explicitly the
dependence on $\log{(1-x')}$ as
\bea
H(1,-1,0;- x' - i 0^{+}) &=& H(0,1,-1;x') + H(1,0;x')H(-1;x') - H(1,-1;x')H(0;x') \nn\\
& & 
- i \pi \left[ H(-1;x')H(1;x') - H(1,-1;x') \right] \, , \nn\\
& = & H(0,1,-1;x') + H(1,0;x')H(-1;x') - H(1,-1;x')H(0;x') \nn\\
& &  - \pi^2
- i \pi \left[ H(-1;x') \log{(x'-1)} - H(1,-1;x') \right] \, ,
\label{ex>1}
\eea
Every function in Eq.~(\ref{ex>1}) is real for $x'>1$ and the imaginary part is
explicitly given by the last term.

\subsubsection{Transformations of Variables and Numerical Evaluation}

A fast and precise numerical evaluation of the HPLs, for all the values of the real 
variable $x$, can be done using an appropriate Taylor expansion in the vicinity of a
point of analyticity of the functions. The strategy is the following (see for
instance  \cite{NUMHPL}):
\begin{enumerate}
\item We focus on the point $x=0$. We extract the possible logarithmic behaviour, 
$\log^n(x)$, of the HPL using the shuffle algebra (we move the rightmost zeroes to the
left). The HPLs with no zeroes on the right are analytic in $x=0$. Correspondingly,
each  HPL takes the form $\sum_{n,m} P_m(x) \log^n(x)$, where $P_m$ is a polynomial of
degree $m$. In the case $x \to 0^-$, the imaginary part comes from $\log^n(x)$, using
the prescription  of Eq.~(\ref{imxg1}).
\item With an appropriate number of terms in the Taylor expansion, one is able to
evaluate numerically the HPL in an interval around $x=0$ with a given precision. In
\cite{NUMHPL} the interval $-(\sqrt{2}-1) \leq x \leq (\sqrt{2}-1)$ is taken as the
central region and using Bernoulli numbers and Chebyshev polynomials, the authors
evaluate the HPLs in double precision using only few terms in the expansion.
\item Using the properties of the HPLs, one can find suitable transformation formulas
for  the argument in order to map different domains of the real axis back to the
central value $-(\sqrt{2}-1) \leq x \leq (\sqrt{2}-1)$. In so doing, using the
formulas found for that region, one is able to cover all the possible values of the
variable $x \in \mathbb{R}$.
\end{enumerate}

\subsection{Generalized Harmonic Polylogarithms \label{GHPL}}

In some physically relevant cases, it can happen that the weight functions defined in
Eq.~(\ref{basicHPLs}) are not sufficient to describe the analytic structure of the
result. Therefore, additional weights (and/or additional structures) have to be taken 
into account, together with the ones introduced in the last section. This gives rise to 
an enlarged set of functions, called Generalized Harmonic Polylogarithms (GHPLs), which
maintain the structure and properties of the HPLs.

Let us focus, for the moment, on the GHPLs that are involved in the calculation of
the NLO light-fermion electroweak corrections to the cross section of production of 
a Higgs boson in gluon fusion and its decay in two photons\footnote{The more extended 
set introduced in \cite{Aglietti:2004tq} will be discussed in the following
sections.}, as considered in \cite{Aglietti:2004nj}. This is a restricted set, that 
contains only four weights, denominated as follows:
\be
G(w_1,w_2, ..., w_n; x) \, , \quad \mbox{with} \, w_i \in \{ -r,-4,-1,0 \} 
\, ,
\ee
according to the definitions given below.
Let furthermore restrict the analysis to the case of real variable  $x \in \mathbb{R}$
(the case of complex $x$ will be treated in section  \ref{complex}).
Therefore, we consider the following set of weight functions:
\bea
g(-r;x) & = & \frac{1}{\sqrt{x(x+4)}} \, , 
\label{w1} \\
g(w;x) & = & \frac{1}{x-w} \, , \quad \mbox{with} \, 
w \in \{ -4,-1,0 \} \, .
\label{w2} 
\eea
These functions have an integrable singularity in $x=0$ and $x=-4$, and  a
non-integrable singularity in $x=w$, respectively.
The related GHPLs of weight 1 are
\bea
G(0;x) & = & \log{(x)} \, , 
\label{w1GHPL0} \\
G(-r;x) & = & \int_{0}^{x} \frac{dt}{\sqrt{t(t+4)}} = - \log{\left( 
\frac{\sqrt{x+4}-\sqrt{x}}{\sqrt{x+4}+\sqrt{x}} \right)} \, , 
\label{w1GHPLmr} \\
G(w;x) & = & \int_{0}^{x} \frac{dt}{t-w} = \log{(x-w)}-\log{(-w)} \, , 
\quad \mbox{with} \, w \in \{ -4,-1 \} \, ,
\label{w1GHPLmw} 
\eea
and they have at most a logarithmic singularity in $x=0$, $x=-1,-4$. $G(0;x)$ and
$G(w;x)$ have a branch cut for $x \leq 0$ and $x \leq w$, respectively. For these
negative values of $x$, they become complex, with imaginary part depending on the
prescription of approach to the cut. $G(-r;x)$ has a branch cut for $x \leq 0$; it 
is purely imaginary in the range $-4 \leq x <0$ and it is a complex number, with 
non vanishing real part, for $x < -4$.

The GHPLs with weight 2 or bigger are defined as repeated integrations of the  weight
functions in Eqs.~(\ref{w1},\ref{w2}):
\be
G(a,{\mathbf w};x) = \int_{0}^{x} dt \, g(a;t) \, G({\mathbf w};t) \, ,
\label{ghpls}
\ee
with the exception of $G({\mathbf 0}_w;x)$, defined as:
\be
G({\mathbf 0}_w;x) = \frac{1}{w!} \log^{w}{(x)} \, .
\label{ghpls0}
\ee
Such a set of functions obeys (by construction) the shuffle algebra of 
Eq.~(\ref{shuffle}) and all the important properties of the HPLs. 

The analytic properties of the functions defined in Eqs.~(\ref{ghpls},\ref{ghpls0}) 
derive from the properties of the logarithm and of the square root.

\subsubsection{Analytic Continuation}

When $x \geq 0$, every GHPL belonging to the set considered in the previous section
is  real and the only possible divergence is a logarithmic divergence  in $x=0$. 

For negative $x$, since the logarithm and the square root have a branch cut  for
negative argument, we must choose how to approach the cut. In order to do that, we
give a vanishing imaginary part to the variable $x$. Let us choose the  following
prescription:
\be
x \to - x' - i \, 0^+ \, .
\ee

The region $x<0$ is divided in three sets, depending on the value of $x'$.  The
analytic continuation has to be done in each region differently. 
\begin{enumerate}
\item For $-1 \leq x < 0$, the imaginary parts come from $\log{(x)}$ and from
the square root, that becomes purely imaginary:
\bea
\log{(x)} & \to & \log{(-x'-i \, 0^+)} = \log{(x')} - i \pi \, , \\
\frac{1}{\sqrt{x(x+4)}} & \to & \frac{1}{\sqrt{(- x' - i \, 0^+)(- x' 
- i \, 0^+ +4)}} = \frac{i}{\sqrt{x'(4-x')}} \, .
\eea
\item For $-4 \leq x < -1$, also the $\log{(x+1)}$ gives an imaginary part:
\be
\log{(x+1)}  \to  \log{(-x'-i \, 0^+ +1)} = \log{(x'-1)} - i \pi \, .
\ee
\item For $- \infty \leq x < -4$, an additional imaginary part comes from the
logarithm $\log{(x+4)}$, while the square root becomes 
real again:
\bea
\log{(x+4)} & \to & \log{(-x'-i \, 0^+ +4)} = \log{(x'-4)} - i \pi \, , \\
\frac{i}{\sqrt{x'(4-x')}} & \to & \frac{i}{\sqrt{(x' + i \, 0^+)(- x' 
- i \, 0^+ +4)}} = - \frac{1}{\sqrt{x'(x'-4)}} \, .
\eea
\end{enumerate}

\subsubsection{Numerical evaluation}

The numerical evaluation of the GHPLs considered in the previous section can be done
in principle using  basically the same strategy of the HPLs.  One focuses on $x=0$,
extracts the logarithmic  behaviour using the shuffle algebra and then expands the
remaining analytic functions (with no zeroes in the rightmost weight).  Then, using
suitable transformations, one relates the basic interval around $x=0$ to the rest of
the real axis. 

However, the actual implementation of this strategy is quite cumbersome. Instead, 
it turns out to be convenient to transform from the beginning the GHPLs with 
square roots in the weights  into a combination of GHPLs with linear weights  using a 
set of variable transformations that will be discussed in the following section. The
advantage of doing so lies in the fact that there exist fast and precise public 
numerical routines, that allow for the evaluation  of generalized
polylogarithms with generic linear weights \cite{Weinzierl}. The latter can be
used to evaluate the GHPLs  belonging to the set discussed in the last section, or,
more in general, to the wider   set introduced in \cite{Aglietti:2004tq}.

\section{Linearization \label{lin}}

The presence, in an analytic result, of GHPLs with square roots together with  linear
weights, is due to the structure of the thresholds and pseudo-thresholds of   the
corresponding Feynman diagrams. 

Let us consider, for instance, the QED corrections to the vertex   diagrams
representing the decay of a photon into an  electron-positron pair. The   particle
content reduces to massless photons and massive electrons/positrons. The   threshold
for the production of the electron-positron pair is at $s = 4 m^2$, while  the
pseudo-threshold lies at $s=0$. Let us look at the differential equations with  
respect to $s$, for the solution of the corresponding MIs. The structure of the 
thresholds and pseudo-thresholds emerges in the homogeneous part with terms such as 
$1/s$ and $1/(s-4m^2)$, that are also present in the  non-homogeneous part (see for
instance \cite{Bonciani:2003te}). The solution  of the homogeneous equation contains
the inverse square root $1/\sqrt{s(s-4m^2)}$. The  particular solution, then, comes
from repeated integrations of $1/\sqrt{s(s-4m^2)}$,  $1/s$, and $1/(s-4m^2)$ terms. In
the Euclidean region ($p^2 = -s >0$) the solution   can be expressed in terms of GHPLs
of the variable $x=p^2/m^2=-s/m^2$, with weights   $-r$, $-4$, and $0$. 

We can get rid of the square root (the weight $-r$) using the following quadratic 
transformation of variable:
\be
x = \frac{(1-\xi)^2}{\xi} \, , \quad 
\xi = \frac{\sqrt{x+4}-\sqrt{x}}{\sqrt{x+4}+\sqrt{x}} \, ,
\label{redvar}
\ee
where $\xi \in \mathbb{C}, \, | \xi | \leq 1$, while $x \in \mathbb{R}$. In fact, we
have:
\bea
\frac{1}{(x+4)} & = & \frac{\xi}{(\xi+1)^2} \, , \\
\frac{1}{\sqrt{x(x+4)}} & = &  - \frac{\xi}{(\xi+1)(\xi-1)} \, .
\eea
Moving from $x$ to $\xi$, the integration measure changes as follows:
\be
\int_0^x dt = \int_1^{\xi} \frac{(\eta+1)(\eta-1)}{\eta^2} \, d \eta \, ,
\label{intmeasure}
\ee
and every GHPLs reduces to a combination of repeated integrations of the simpler weight 
functions defined in Eqs.~(\ref{basicHPLs}). As a consequence of that, the set of 
weights $\{ -r,-4,0 \}$ is transformed into the set $\{ -1,0,1 \}$, and the GHPLs are
transformed into the usual HPLs defined in section \ref{HPLs}.

Let us consider, now, a more complicated problem, in which zero- and multiple-mass cuts 
are present at the same time. This is, for instance, the case of the electroweak 
corrections to leptonic or hadronic processes in which the lepton and quark masses are 
neglected and only the vector boson masses are considered different from zero.
In this case, the homogeneous part of the differential equations for the
corresponding MIs contain terms as $1/s$, $1/(s-4m^2)$, $1/\sqrt{s(s-4m^2)}$, 
together with terms as $1/(s-m^2)$. Therefore, the weights $-r$, $-4$, $-1$, and $0$ 
are present at the same time.
In this situation, it is more difficult to get rid of the square root. If we require, 
for instance, that the weights belong always to the set of real numbers, 
$w_i \in \mathbb{R}$, there is no transformation of variable that could linearize all
the weights $\{ -r, -4, -1, 0 \}$ at the same time. 
However, if we relax this constraint, we can move from the set with square  roots and
linear weights to a set of only linear weights, using the change of variable
(\ref{redvar}).

Using Eq.~(\ref{redvar}), the old weight functions are transformed into:
\bea
g(-r;t) & = & \frac{1}{\sqrt{t(t+4)}} = - \frac{\eta}{(\eta+1)(\eta-1)}  \, ,
\label{ch1} \\
g(-4;t) & = & \frac{1}{t+4} = \frac{\eta}{(\eta+1)^2} \label{ch2} \, , \\
g(-1;t) & = & \frac{1}{t+1} = \frac{\eta}{(\eta-c)(\eta-\bar{c})} \, , 
\label{ch3} \\
g(0;t) & = & \frac{1}{t} = \frac{\eta}{(\eta-1)^2} \, ,
\label{ch4}
\eea
with
\be
c  =  \frac{1+i \sqrt{3}}{2} = e^{i \frac{\pi}{3}} \, , \quad
\bar{c}  =  \frac{1-i \sqrt{3}}{2} = e^{-i \frac{\pi}{3}}  \, ,
\label{zeta}
\ee
where $c$ and $\bar{c}$ are the two primitive sixth roots of the unity.
Then, combining the integration measure, Eq.~(\ref{intmeasure}), with the
Eqs.~(\ref{ch1}--\ref{ch4}), the original GHPLs with square root in the weight are
transformed into 
\bea
G(-r,{\mathbf w};x) & = & \int_0^x dt \, g(-r;t) \, G({\mathbf w};t) 
= - \int_1^{\xi} d \eta \, \frac{1}{\eta} \, G({\mathbf w};t(\eta)) \, ,
\label{int1} \\
G(-4,{\mathbf w};x) & = & \int_0^x dt \, g(-4;t) \, G({\mathbf w};t) 
= \int_1^{\xi} d \eta \, \left( - \frac{1}{\eta} + \frac{2}{\eta+1} \right) 
\, G({\mathbf w};t(\eta)) \, ,
\label{int2} \\
G(-1,{\mathbf w};x) & = & \int_0^x dt \, g(-1;t) \, G({\mathbf w};t)  = \nn\\
& = &
\int_1^{\xi} d \eta \, \left( - \frac{1}{\eta} + \frac{1}{\eta-c} 
+ \frac{1}{\eta-\bar{c}} \right) \, G({\mathbf w};t(\eta)) \, ,
\label{int3} \\
G(0,{\mathbf w};x) & = & \int_0^x dt \, g(0;t) \, G({\mathbf w};t) 
= \int_1^{\xi} d \eta \, \left( - \frac{1}{\eta} + \frac{2}{\eta-1} \right) 
\, G({\mathbf w};t(\eta)) \, .
\label{int4} 
\eea
Therefore, the set $\{ -r,-4,-1,0 \}$ of weights with square roots,  has been
transformed into a new set, with only linear weights: $\{ -1,0,1,c,\bar{c} \}$. 
This new set, contains the original HPLs, discussed in section \ref{HPLs}, and 
new GHPLs with complex weights $c$ and $\bar{c}$. The latter, have branch cuts 
in the complex $x$ plain, starting at $x=c,\bar{c}$ respectively. At weight 1, 
they are:
\bea
G(c;x) & = & \int_0^x \frac{dt}{t-c} = \log{(x-c)} - \log{(-c)} \, , \\
G(\bar{c};x) & = & \int_0^x \frac{dt}{t-\bar{c}} = \log{(x-\bar{c})} 
- \log{(-\bar{c})} \, .
\eea

We can summarize the linearization procedure as follows:
\begin{enumerate}
\item We base our analysis in the region in which $x \geq 0$ (the formulas will be
afterwards analytically continued in the region $x<0$, if necessary).
\item We transform the integration variable in the new variable $\eta$, on which we
integrate from 1 to $\xi$. Troubles with the integration in $\eta=1$ can occur, due  to
possible singular behaviours. Since such singularities can occur only from the weights
0 in the variable $x$, we avoid the possible logarithmic divergence in $\eta=1$ using
the shuffle  algebra and extracting the trailing zeroes in $x$. The logarithms so
found,  $G({\mathbf 0}_n;x)$, are directly rewritten as $1/n! \, \log^n{(x)}$ and,
then, straightforwardly transformed in the variable $\xi$ using the relation $\log{(x)}
= 2 \log{(1-\xi)} - \log{(\xi)}$.
\item We linearize the GHPLs of weight 1.
\item Weight-by-weight we proceed to the linearization of the GHPLs with weight
2 and bigger, integrating over the new integration measure the corresponding
linearized GHPL times the corresponding linearized weight function. 
\end{enumerate}

As an example, we give here the expressions of the linearized GHPLs with weight 1.

$G(0;x)$ can be converted directly in the new variable $\xi$, since\footnote{
Note again the different sign with respect to the weight $+1$ in the
Remiddi-Vermaseren notation \cite{HPLs}.
}:
\be
G(0;x) = \log{(x)} = 2 \log{(1-\xi)} - \log{(\xi)} = 2 G(1;\xi) - G(0;\xi) \, .
\ee
Using the relations in Eqs.~(\ref{int1}--\ref{int4}), we have:
\bea
G(-r;x) & = & - \int_1^{\xi} \frac{d \eta }{\eta} = - G(0;\xi) \, , \\
G(-4;x) & = & \int_1^{\xi} d \eta \, \left( - \frac{1}{\eta} 
+ \frac{2}{\eta+1} \right) = - 2 \log{(2)} + 2 G(-1;\xi) - G(0;\xi) \, , \\
G(-1;x) & = & \int_1^{\xi} d \eta \, \left( - \frac{1}{\eta} + \frac{1}{\eta-c} 
+ \frac{1}{\eta-\bar{c}} \right) \, , \nn\\
& = & 
       - G(c;1)
       - G(\bar{c};1)
       + G(c;\xi)
       + G(\bar{c};\xi)
       - G(0;\xi) \, .
\eea

It is worth to notice that the linearization algorithm generates some constants, {\it
i.e.} the linearized GHPLs evaluated in $\xi=1$. In many cases, these constants have a
representation in terms of known transcendental constants. In general, however, this
is not true. For the purpose of the numerical evaluation  of the GHPLs with square
roots in the weights using existing {\tt C++} routines, these constants can be left as
they are. In fact, the routines provide a fast and accurate numerical evaluation in
every point, and then also in  $\xi=1$. In our particular case we have\footnote{The
constants (and also the GHPLs with complex weights in the  actual expressions for the
MIs) should appear always in a way such that their sum is real,  as it has to be,
since we are in the Euclidean region.}:
\be
G(c;1) + G(\bar{c};1) = 0 \, ,
\ee
such that
\be
G(-1;x) = G(c;\xi) + G(\bar{c};\xi) - G(0;\xi) \, .
\ee

Knowing the expressions of the linearized GHPLs with weight 1 and linearized 
weight functions, we can proceed with the linearization of the GHPLs at weight 2. 
If we consider, for instance, the function $G(0,-1;x)$ we have:
\bea
G(0,-1;x) & = & \int_0^x \frac{dt}{t} G(-1;t) \, , \nn\\
& = & \int_1^{\xi} d \eta \, \left( - \frac{1}{\eta} + \frac{2}{\eta-1} \right) 
\, \left[ G(c;\eta) + G(\bar{c};\eta) - G(0;\eta) \right] \, , \nn\\
& = & \zeta(2)
          - G(0,\bar{c};\xi)
          - G(0,c;\xi)
          + \frac{1}{2} G(0;\xi)^2
          - 2 G(1;\xi) G(0;\xi)\nn\\
& &        
          + 2 G(0,1;\xi) 
	  + 2 G(1,\bar{c};\xi)
          + 2 G(1,c;\xi)
\, .
\eea
In the same way one can proceed for higher weights. Explicit formulas for the weight-2
and weight-3 GHPLs involved in the NLO electroweak  corrections for the production of
a Higgs boson in gluon fusion are provided in appendix \ref{W2APP} and appendix
\ref{W3APP}, respectively.

\subsection{Analytic Continuation of the Linearized GHPLs and their Numerical
Evaluation \label{anconlin}}

The analytic continuation of the linearized GHPLs is less complicated than the  one
concerning their original form. In fact, while the variable $x$ ranges from $\infty$
to $0$, the corresponding variable $\xi$ is real and positive and it ranges from $0$
to $1$. When $x$ becomes negative, but in the range  $-4 \leq x < 0$, 
\be
x \to - x' - i \, 0^+ \, , \quad 0 < x' \leq 4 \, ,
\ee
$\xi$ becomes imaginary:
\be
\xi = \frac{\sqrt{x+4}-\sqrt{x}}{\sqrt{x+4}+\sqrt{x}} \, \, \to \, \, 
\zeta = \frac{\sqrt{4-x'}+i\sqrt{x'}}{\sqrt{4-x'}-i\sqrt{x'}} = e^{i \, 2 \phi} \, ,
\ee
where
\be
\phi = \arctan{\sqrt{ \frac{x'}{4-x'} }} \, , \quad 0 < \phi \leq \frac{\pi}{2} 
\, .
\ee
Finally, when $x'$ ranges from $4$ to $\infty$, $\xi$ becomes real again:
\be
\zeta \, \, \to \, \, 
\xi' = \frac{\sqrt{x'}-\sqrt{x'-4}}{\sqrt{x'}+\sqrt{x'-4}}  \, ,
\ee
and it ranges from $1$ to $0$.
We must, therefore, discuss three regions.
\begin{enumerate}
\item For $0 \leq x < \infty$, we have $0 < \xi < 1$. The original GHPLs are
real. The linearized GHPLs contain functions that are manifestly real, as the
ones with weights $-1,0,1$, but also functions that are complex: those that
contain the weights $c$ and $\bar{c}$. However, the GHPLs containing the weights 
$c$ and $\bar{c}$ appear in the formulas always in pairs (for instance
$G(-1,c;\xi)+G(-1,\bar{c};\xi)$), in such a way that,
although the single GHPLs of the pair are complex, their sum is real, since the imaginary
parts are equal and opposite. 
The numerical evaluation of such GHPLs can be done straightforwardly using the
VW routines presented in \cite{Weinzierl}.
\item For $-4 \leq x < 0$, $\xi$ is a pure phase, $\xi = e^{i \, 2 \phi}$.
In order to evaluate numerically the GHPLs in this region, 
we have to notice that the VW routines,
while allowing the use of complex weights,
do not provide the possibility of evaluation of GHPLs with complex argument. 
However, for the GHPLs with non-trailing zeroes the following general formula
holds:
\be
G(w_1,w_2,...,w_n;x) = G(\lambda w_1,\lambda w_2,...,\lambda w_n;\lambda x) \, ,
\quad \lambda \in \mathbb{C} \, ,
\ee
as it will be discussed in the next section.
Extracting the trailing zeroes and then choosing\footnote{Actually, it is sufficient to
choose $\lambda = e^{- i \, arg(x)}$.}
\be
\lambda = \frac{1}{x} \, ,
\ee
the GHPLs under consideration are transformed in GHPLs of real
argument, $\xi=1$, and complex weights,
$\{ \pm e^{-i \, 2 \phi}, 0, e^{-i \, \left( 2 \phi \, \pm \frac{\pi}{3}\right) } \}$,
that can be evaluated using again the VW routines.
\item For $-\infty < x < -4$, $\xi$ is again real and we are back to the case
explained in the first point.
\end{enumerate}

\section{GHPLs of a Complex Variable \label{complex}}

In this section, we consider the case in which the GHPLs have to be evaluated  in the
complex plain. Therefore, $x$ is complex from the beginning\footnote{The case in
which $x$ is real, but the corresponding reduced variable $\xi$ is complex, was
already discussed in the previous section. }. We are particularly interested in the
following situation. Let us suppose that the dimensionless variable $x$ is indeed a
ratio between two physically meaningful variables: a squared momentum and a squared
mass:
\be
x = \frac{p^2}{m^2} = - \frac{s}{m^2} \, ,
\ee 
with $\sqrt{s}$ the c.m. energy of a certain process. This is,  for instance, the case
of the corrections presented in \cite{Aglietti:2004nj}, but it is a quite general
assumption. If the particle to which the mass $m$ belongs is an unstable particle, its
width $\Gamma$ is going to play an active  role in the determination of the
corresponding physical observable.  Consequently, the parameter $x$ becomes complex,
since we should now consider
\be
x = - \frac{s}{(m-i \Gamma /2)^2} = - \frac{s}{M^2} e^{i \phi} \, ,
\label{xcompl}
\ee
where
\be
M^2 = m^2 - \frac{\Gamma^2}{4} \, , \quad \mbox{and} \quad
\phi = \arctan{\left\{ \frac{m \Gamma}{m^2- \frac{\Gamma^2}{4}} \right\}}
\, .
\ee
In the non-physical region, in which $s<0$, we have from Eq.~(\ref{xcompl}) that
\be
{\mathcal Re} (x) > 0 \, , \quad \mbox{and} \quad {\mathcal Im} (x) > 0 \, .
\label{fquad}
\ee
The variable $\xi$ defined in Eq.~(\ref{redvar}), correspondingly, is also complex.
The definition of the GHPLs does not change, except from the fact that now the
integration is over a curve in the complex plain. Since the functions are analytic in
the region defined by Eq.~(\ref{fquad}), the value of the GHPL does not depend on the
path.
In this region we have:
\be
\xi = \frac{\sqrt{x+4}-\sqrt{x}}{\sqrt{x+4}+\sqrt{x}} = 
\frac{\sqrt{r_1+x_1+4} - \sqrt{r_2+x_1} + i \, \bigl( \sqrt{r_1-x_1-4}
- \sqrt{r_2-x_1} \bigr)}{\sqrt{r_1+x_1+4} + \sqrt{r_2+x_1} 
+ i \, \bigl( \sqrt{r_1-x_1-4} + \sqrt{r_2-x_1} \bigr) } \, ,
\ee
where:
\bea
x_1 & = & {\mathcal Re} (x) > 0 \, , \\
x_2 & = & {\mathcal Im} (x) > 0 \, , \\
r_1 & = & \sqrt{(x_1+4)^2+x_2^2} \, , \\
r_2 & = & \sqrt{x_1^2+x_2^2} \, .
\eea
Therefore, we are in the situation in which we have to evaluate GHPLs with linear
complex weights as functions of a complex variable $\xi$.

Let us consider a generic GHPL, $G(w_1,w_2,...,w_n;x)$ in the case in which  $w_i,x
\in \mathbb{R}$. If no trailing zeroes  are present, we can define a non-vanishing
real parameter $\lambda \in \mathbb{R}$, such that the following scale invariance
holds:
\be
G(w_1,w_2,...,w_n;x) = G(\lambda w_1,\lambda w_2,...,\lambda w_n;\lambda x) \, .
\label{scale}
\ee

The demonstration of Eq.~(\ref{scale}) can be done by induction. It is trivially
verified for $n=1$ ($\lambda,w_1 \not = 0$). In fact:
\be
G(\lambda w_1;\lambda x) = \int_0^{\lambda x} dt \, g(\lambda w_1;t) \, ,
\ee
and moving to the new integration variable $r = t / \lambda$, we have:
\be
G(\lambda w_1;\lambda x) = \int_0^{x} \lambda dr \, g(\lambda w_1; \lambda r)
= \int_0^{x} dr \, g(w_1; r) = G(w_1;x) \, .
\ee
Let us suppose it is verified for $n=i$. For $n=i+1$ we have:
\bea
G(\lambda w_{i+1}, \lambda \mathbf{w};\lambda x) &=& 
\int_0^{\lambda x} dt \, g(\lambda w_{i+1};t) 
G(\lambda \mathbf{w};t) =
\int_0^{x} \lambda dr \, g(\lambda w_1; \lambda r) 
G(\lambda \mathbf{w}; \lambda r) \, , \nn\\
& = & \int_0^{x} dr \, g(w_1; r)G(\mathbf{w}; r) =
G(w_{i+1}, \mathbf{w}; x) \, .
\eea

Let us suppose, now, that $w_i,\lambda,x \in \mathbb{C}$. For the weight 1 we have
(remember that we are considering the case in which $|\lambda|,|w_1| \not = 0$):
\be
G(\lambda w_1;\lambda x) = \int_{0,\gamma}^{\lambda x} dz \, g(\lambda w_1;z) =
\int_{0,\gamma}^{\lambda x} \frac{dz}{z-\lambda w_1} \, ,
\ee
where $\gamma$ is a path in the complex plain connecting the origin, $z=0$, to the
point $\lambda x = |\lambda| |x| e^{i(arg(\lambda)+arg(x))} = |\lambda| |x| 
e^{i(\Lambda+X)}$. If we rescale the integration variable by the real number 
$|\lambda| |x|$, we have
\be
G(\lambda w_1;\lambda x) = \int_{0,\gamma'}^{e^{i(\Lambda+X)}} 
\frac{dz'}{z'- \xi e^{i(\Lambda+W_1)}} \, ,
\ee
where $\xi = |w_1|/|x|$ and $W_1 = arg(w_1)$. The path $\gamma'$ connects the origin
and the point on the circle of radius 1 with argument $(\Lambda+X)$. Let us define 
$\gamma_1$ the path along the radius from the origin to $e^{i(\Lambda+X)}$.
$\Gamma=\gamma'-\gamma_1$ is a closed path that we suppose not to include the pole
$z'=\xi e^{i(\Lambda+W_1)}$. The integral along the path $\Gamma$ vanishes for the
Cauchy's theorem. 
The integral over the radius can be rewritten as a one-dimensional
integral of real variable with the substitution
$t=z'\exp\left( -i(\Lambda+X) \right)  $.
Therefore:
\be
G(\lambda w_1;\lambda x) = \int_{0,\gamma_1}^{e^{i(\Lambda+X)}} 
\frac{dz'}{z'- \xi e^{i(\Lambda+W_1)}} = \int_{0}^{1}
\frac{dt}{t- \xi e^{i(W_1-X)}} = G(w_1/x;1) \, .
\ee
On the other hand, we have also:
\be
G(w_1; x) = \int_{0,\gamma_1}^{e^{i(X)}} 
\frac{dz'}{z'- \xi e^{i(W_1)}} = \int_{0}^{1}
\frac{dt}{t- \xi e^{i(W_1-X)}} = G(w_1/x;1) \, ,
\label{resc1}
\ee
thus,
\be
G(\lambda w_1;\lambda x) = G(w_1; x) \, .
\ee
Note that, in the end, for our purposes, we can just use Eq.~(\ref{resc1}).

Let us suppose, now, that the rescaling is verified for $n=i$. For $n=i+1$ we have:
\bea
G(\lambda w_{i+1}, \lambda \mathbf{w};\lambda x) \! \! &=& \! \! 
\int_{0,\gamma_1}^{\lambda x} dz \, g(\lambda w_{i+1};z) \,
G(\lambda \mathbf{w};z) = \int_{0,\gamma_1}^{e^{i(\Lambda+X)}} \hspace*{-8mm}
\frac{dz'}{z'- \xi e^{i(\Lambda+W_{i+1})}} 
G(\lambda \mathbf{w}; |\lambda | |x| z') \, , \nn\\
\! \! &=& \! \! \int_{0}^{1} \frac{dt}{t- \xi e^{i(W_{i+1}-X)}} \, 
G(\lambda \mathbf{w}; \lambda x \, t) \, , \nn\\
\! \! &=& \! \! \int_{0}^{1} \frac{dt}{t- \xi e^{i(W_{i+1}-X)}} \, 
G(\mathbf{w}/x; t) \, = G(w_{i+1}/x,\mathbf{w}/x; 1) \, .
\label{complip1}
\eea
Choosing $\lambda=1$ in Eq.~(\ref{complip1}), we can demonstrate that
\be
G(w_{i+1}, \mathbf{w}; x) = G(w_{i+1}/x,\mathbf{w}/x; 1) \, ,
\label{rescN}
\ee
and, therefore
\be
G(\lambda w_{i+1}, \lambda \mathbf{w};\lambda x)  = G(w_{i+1}, \mathbf{w}; x) \, .
\ee

Using Eq.~(\ref{rescN}), we can employ the numerical routines provided in 
\cite{Weinzierl} for the evaluation of the GHPLs. In fact, now the GHPLs have complex 
weights (ratios of the original weights $w_i$ and the variable $x$), but real 
variable\footnote{
Note that it is sufficient to divide by $e^{i \, arg(x)}$}, equal to 1.

Let us consider again our set $\{ -r,-4,-1,0 \}$, and see what happens in the 
different regions.
The analytic continuation from the non-physical $s<0$ region to the physical
region in which $p^2 \to - s - i\, 0^+$, with $s>0$, corresponds to the
transformation $x \to - x'$, where, now, $x' \in \mathbb{C}$ and it is defined 
as follows:
\be
x' = x_1' + i \, x_2' = \frac{s}{M^2} e^{i \phi} \, .
\ee
Correspondingly, the variable $\xi$ becomes $\xi \to \zeta$, with $\zeta \in
\mathbb{C}$ defined as follows:
\be
\zeta = \frac{\sqrt{4-x'}-\sqrt{-x'}}{\sqrt{4-x'}+\sqrt{-x'}} = 
\frac{\sqrt{r_1'-x_1'+4} - \sqrt{r_2-x_1'} - i \, \bigl( \sqrt{r_1'+x_1'-4}
- \sqrt{r_2+x_1} \bigr)}{\sqrt{r_1'-x_1'+4} + \sqrt{r_2-x_1'} - i \, 
\bigl( \sqrt{r_1'+x_1'-4} + \sqrt{r_2+x_1} \bigr)} \, ,
\ee
where now
\be
r_1' = \sqrt{(4-x_1)^2+x_2^2} \, .
\ee
Note that $\zeta$ does not have anymore modulus 1, as it was the case of real
$x$ shown in Eq.~(\ref{zeta}).

\section{Generalizations and Additional Weights \label{general}}

In this section, we enlarge the set of possible weights in order to cover  
the GHPLs needed for the analytic expressions of the MIs in \cite{Aglietti:2004tq}. 
The goal is to be able to describe the following set:
\be
\{ -1-r,-r,-4,-1,0,1,4,r,1+r,c,\bar{c} \} \, ,
\label{completeset}
\ee
where the additional weight functions (not introduced in the previous sections) are defined 
as follows\footnote{Note the difference in sign in the definition of $g(4;x)$, $g(r;x)$, 
and $g(1+r;x)$ with respect to \cite{Aglietti:2004tq}.}:
\bea
g(4;x) & = & \frac{1}{x-4} \, , \\
g(r;x) & = & \frac{1}{\sqrt{x(x-4)}} \, , \\
g(1+r;x) & = & \frac{1}{\sqrt{x(x-4)}(x-1)} \, , \\
g(-1-r;x) & = & \frac{1}{\sqrt{x(x+4)}(x+1)} \, . 
\eea
The guidelines sketched in this section can be used for other, more complicated, sets.

It is first worth to notice that the possible weights listed in 
Eq.~(\ref{completeset}) do not appear all together at the same time. The 
appearance of a particular weight in a GHPL depends on the cut structure of the 
relative Feynman diagram. In the MIs presented in \cite{Aglietti:2004tq} 
we cannot have, for instance, the weights $r$ and $-r$ at the same time in the same 
GHPL. The same happens for the pair $(c,\bar{c})$ with the square roots $r$ or $-r$. 
Actually, the structure of the MIs in \cite{Aglietti:2004tq} is such that we are 
concerned effectively with three different subsets, that form each a closed base. 
They are:
\be
\{ -1,0,1,c,\bar{c} \} \, , \quad \{ -1-r, -r,-4,-1,0 \} \, , \quad \{ 0,1,4,r,1+r \} \, .
\ee
The three subsets do not mix with each other and they can be linearized (once and for
all) using different variable transformations.

\subsection{The set $\mathbf \{ -1,0,1,c,\bar{c} \}$}

The GHPLs belonging to this set can be evaluated straightforwardly with the help 
of the routines in \cite{Weinzierl} without any further variable transformation. In 
the case in which the variable $x$ is complex, we just have to use the scale 
invariance of the GHPLs, as explained in section \ref{complex}.

\subsection{The set $\mathbf \{ -1-r,-r,-4,-1,0 \}$}

This set contains the weights treated in section \ref{lin}, $\{ -r,-4,-1,0 \}$, 
with a small enlargement due to the weight $(-1-r)$. Note that this enlargement 
is totally painless, since the new weight $(-1-r)$ transforms in the same set of 
linearized weights $\{ -1,0,1,c,\bar{c} \}$. In fact,
\bea
G(-1-r,{\mathbf w};x) & = & \int_0^x dt \, g(-1-r;t) \, G({\mathbf w};t) 
= \int_0^x \frac{dt}{\sqrt{t(t+4)} (t+1)} \, G({\mathbf w};t) \, , \nn\\
& = & i \frac{\sqrt{3}}{3} \int_1^{\xi} d \eta \, 
\left( \frac{1}{\eta-c} - \frac{1}{\eta-\bar{c}} \right) \, 
G({\mathbf w};t(\eta)) \, .
\label{mrm1}
\eea
Note that the GHPL $G(-1-r,{\mathbf w};x)$, which is real for $x \geq 0$, is
written as a difference of the two complex GHPLs: $G(c,...;\eta)$ and 
$G(\bar{c},...;\eta)$. This difference is indeed complex, since the two GHPLs
have the same real part but opposite imaginary parts. The factorized $i$ in
Eq.~(\ref{mrm1}) makes in such a way that the combination is real.

\subsection{The set $\mathbf \{ 0,1,4,r,1+r \}$}

These positive weights cannot be linearized with the change of variable in 
Eq.~(\ref{redvar}). 
Instead, we must use the change of variable that was used in 
\cite{Bonciani:2003hc}:
\be
x = \frac{(1+\omega)^2}{\omega} \, , \quad
\omega = \frac{\sqrt{x}-\sqrt{x-4}}{\sqrt{x}+\sqrt{x-4}} \, .
\label{redvarp}
\ee
When $x$ is positive and ranges from $\infty$ to $4$, the corresponding variable
$\omega$ ranges between 0 and 1. When $0\leq x < 4$, $\omega$ becomes imaginary.
Giving to $x$ a negative vanishing imaginary part (anticipating the prescription for 
the continuation to the Minkowski region), we have:
\be
\omega = \frac{\sqrt{x}-\sqrt{x-4}}{\sqrt{x}+\sqrt{x-4}} \to
\omega' = \frac{\sqrt{x}-\sqrt{x-4-i0^+}}{\sqrt{x}+\sqrt{x-4-i0^+}} = 
\frac{\sqrt{x}+i\sqrt{4-x}}{\sqrt{x}-i\sqrt{4-x}} = e^{i \, 2\phi} \, ,
\ee
where
\be
\phi = \arctan{\sqrt{\frac{4-x}{x}}} \, , \quad 0 \leq \phi < \frac{\pi}{2} \, .
\ee
Finally, when $x$ becomes negative,
\be
x \to - x' - i 0^+ \, , \quad x' > 0 \, ,
\ee
we have
\be
\omega' \to \omega'' = \frac{\sqrt{x'+4}-\sqrt{x'}}{\sqrt{x'+4}+\sqrt{x'}} \, , 
\ee
and $\omega''$ ranges between 1 and 0 when $x'$ ranges from 0 to $\infty$.

Moving from $x$ to $\omega$, the integration measure changes as follows:
\be
\int_0^x dt = \int_{-1}^{\omega} \frac{(\eta+1)(\eta-1)}{\eta^2} \, d \eta \, .
\label{intmeasure2}
\ee
Using eq.(\ref{redvarp}) the old weight functions 
are transformed into:
\bea
g(0;t) & = & \frac{1}{t} = \frac{\eta}{(\eta+1)^2} \, ,  
\label{dh1} \\
g(1;t) & = & \frac{1}{t-1} = \frac{\eta}{(\eta+c)(\eta+\bar{c})} \, , 
 \label{dh2} \\
g(4;t) & = & \frac{1}{t-4} = \frac{\eta}{(\eta-1)^2} \, , 
\label{dh3} \\
g(r;t) & = & \frac{1}{\sqrt{t(t-4)}} = - \frac{\eta}{(\eta+1)(\eta-1)} \, ,
\label{dh4} \\
g(1+r;t) & = & \frac{1}{\sqrt{t(t-4)}(t-1)} = 
- \frac{\eta^2}{(\eta+1)(\eta-1)(\eta+c)(\eta+\bar{c})} \, ,
\label{dh5}
\eea
where the complex numbers $c$ and $\bar c$ were defined in section \ref{lin}.

Combining Eq.~(\ref{intmeasure2}) with Eqs.~(\ref{dh1}--\ref{dh5}),
we have the following transformation formulas for the definition of the GHPLs:
\bea
\hspace{-5mm} 
G(0,{\mathbf w};x) \hspace{-2mm} & = & \hspace{-3mm} \int_0^x dt \, g(0;t) \, G({\mathbf w};t) 
= \int_{-1}^{\omega} d \eta \, \left( - \frac{1}{\eta} + \frac{2}{\eta+1} \right) 
\, G({\mathbf w};t(\eta)) \, , \\
\hspace{-5mm} 
G(1,{\mathbf w};x) \hspace{-2mm} & = & \hspace{-3mm} \int_0^x dt \, g(1;t) \, G({\mathbf w};t) =
\int_{-1}^{\omega} d \eta \, \left( - \frac{1}{\eta} + \frac{1}{\eta+c} 
+ \frac{1}{\eta+\bar{c}} \right) \, G({\mathbf w};t(\eta)) \, , \\
\hspace{-5mm} 
G(4,{\mathbf w};x) \hspace{-2mm} & = & \hspace{-3mm} \int_0^x dt \, g(4;t) \, G({\mathbf w};t) 
= \int_{-1}^{\omega} d \eta \, \left( - \frac{1}{\eta} + \frac{2}{\eta-1} \right) 
\, G({\mathbf w};t(\eta)) \, , \\
\hspace{-5mm} 
G(r,{\mathbf w};x) \hspace{-2mm} & = & \hspace{-3mm} \int_0^x dt \, g(r;t) \, G({\mathbf w};t) 
= - \int_{-1}^{\omega} d \eta \, \frac{1}{\eta} \, G({\mathbf w};t(\eta)) \, , \\
\hspace{-5mm} 
G(1+r,{\mathbf w};x) \hspace{-2mm} & = & \hspace{-3mm} \int_0^x dt \, g(1+r;t) \, 
G({\mathbf w};t) = 
i \frac{\sqrt{3}}{3} \! \int_{-1}^{\omega} \! d \eta \, \left( \frac{1}{\eta\! +\! \bar{c}} 
- \frac{1}{\eta\! +\! c} \right)  G({\mathbf w};t(\eta))  .
\eea

The integration in $\eta$ deserves a further discussion.
As in the case already presented in section \ref{lin}, the point $\eta=-1$ can 
be source of a non integrable singularity. However, the possible divergence in 
$\eta=-1$ is connected to the original point $x=0$, and then, ultimately, to 
the right-most weight 0 in the GHPLs of $x$. It is sufficient, therefore, to 
extract the right-most trailing zeroes in $x$ before the change of variable 
(\ref{redvarp}) is applied, using the shuffle algebra. The functions 
$G({\mathbf 0}_n;x) = 1/n! \, \log^n{(x)}$ can be directly transformed in the 
new variable $\omega$ using the relation
$\log{(x)} = 2 \log{(\omega+1)} - \log{(\omega)}$.
The GHPLs that do not contain trailing zeroes in the right-most weights are 
regular in $\eta=-1$ after the variable transformation.

\subsection{Mixed Weights}

Although the weights belonging to the different sets described above do not mix
in the expressions of the MIs of \cite{Aglietti:2004tq}, we can further extend
the analysis and try variable transformations that linearize wider sets of weights.
This can be done provided that we do not mix the square roots with different signs.
For instance, it can be shown that the weights belonging to the set 
$\{ -1-r,-r,-4,-1,0,1,4 \}$ can be linearized at the same time, using the 
variable transformation in Eq.~(\ref{redvar}).
Analogously, the set $\{ -4,-1,0,1,4,r,1+r \}$ can be linearized with the help of 
the change of variable of Eq.~(\ref{redvarp}).

\section{Two-loop Light-Fermion contributions to the Higgs Production in Gluon
Fusion \label{higgs}}

In this section, we revisit the calculation of the NLO light-fermion electroweak 
corrections to the Higgs boson production in gluon fusion. 

In \cite{Aglietti:2004nj} these corrections were evaluated analytically,  and the
results were expressed in terms of GHPLs with square root in the weights. The numerical
evaluation was done using real $W$ and $Z$ masses and with FORTRAN routines written {\it
ad hoc}\footnote{
In \cite{Degrassi:2004mx}, the remaining electroweak corrections due to the top 
quark were calculated as a Taylor expansion in $m_H^2/(4m_W^2)$. 
Finally, in \cite{Actis:2008ug} a numerical calculation with complex $W$ and $Z$
masses was done for the complete set of NLO electroweak corrections.
}. The electroweak corrections appear to be very peaked at $m_H\sim 2 m_W$ and
$m_H\sim 2 m_Z$ because of the opening of the two corresponding thresholds. In this
section we recompute the NLO-EW corrections using the VW routines employing complex
values for the $W$ and $Z$ masses. As a result, the finite $W$ and $Z$ widths smear the
peaks at the thresholds and resize the relative importance of the corrections in the
region $m_H \sim 2 m_W, 2 m_Z$.

Neglecting QCD corrections, the partonic production cross section, up to 2-loop level,
has the following form:
\be
\sigma(gg \to H) = \frac{G_F \alpha_S^2}{512 \, \sqrt{2} \, \pi} \, 
|{\mathcal G}^{1l} + \alpha \, {\mathcal G}^{2l}_{EW} |^2 \, ,
\ee
where $G_F$ is the Fermi constant, $\alpha_S$ the strong coupling constant, and
$\alpha$ the fine structure constant.

The lowest order, ${\mathcal G}^{1l}$, is due to one-loop diagrams with heavy quarks
running in the loop. The dominant contribution comes from a loop of top, while the
contribution of a b-quark loop is of the order of some percents of the previous one.
The analytic expression of ${\mathcal G}^{1l}$ is:
\be
{\mathcal G}^{1l} = \sum_{q=t,b}\frac{4}{x_{q}} \left[ 2 - \left( 1 + \frac{4}{x_{q}} 
\right) G(-r,-r;x_{q}) \right] \, ,
\ee
where $x_{q}= - m_H^2/m_{q}^2$, $m_{q}$ is the heavy-quark mass (top or bottom mass),
and the analytic continuation has to be taken considering a  positive vanishing $m_H$
imaginary part: $x_{q} \to - x'_{q}-i\,0^+$,  where $x'_{q} = m_H^2/m_{q}^2$.
$G(-r,-r;x_{q})$ can be immediately transformed into a square logarithm of the variable
$\xi$, defined in  Eq.~(\ref{redvar}), as for instance in Eq.~(\ref{Hmrmr}).

The two-loop electroweak light-fermion contributions, ${\mathcal G}^{2l}_{lf}$, to
${\mathcal G}^{2l}_{EW}$ can be expressed as \cite{Aglietti:2004nj}:
\be
{\mathcal G}^{2l}_{lf} = \frac{(m_W-i \Gamma_W/2)^2}{2 \pi s^2 \, m_H^2} \left[ 
  \frac{2} {c^4} \left( \frac{5}{4} - \frac{7}{3} \, s^2 +
\frac{22}{9} s^4 \right)  A_1 (x_Z) + 
4 \, A_1 (x_W) \right] \, ,
\ee
where $s^2 = \sin^2 \theta_W$, $c^2 = 1- s^2$,
\be
x_W = - \frac{m_H^2}{(m_W-i\Gamma_W/2)^2} \, ,
\quad x_Z = - \frac{m_H^2}{(m_Z-i\Gamma_Z/2)^2}
\ee
and
\bea
A_1 (x) & = & -4 + 2 \left( 1+\frac{1}{x} \right) G(-1;x) 
                 + \frac{2}{x} G(0,-1;x)
		 + 2 \left( 1+\frac{3}{x} \right) G(0,0,-1;x) \nn\\
& &
		 + \left( 1+\frac{2}{x} \right) \left[  2 G(0, - r, - r;x)
		                               - 3 G( - r, - r,-1;x) \right]
		 - \sqrt{x (x+4)} \Biggl\{ 
                      \frac{2}{x} G( - r;x) \nn\\
& &
		      + \frac{x+2}{x^2} \Bigl[  2 G( - r, - r, - r;x)
		         + 2 G( - r,0,-1;x) - 3 G(-4, - r,-1;x) \Bigr]
			           \Biggr\} \, .
\label{A1bis}
\eea
The GHPLs with square root in the weights involved in Eq.~(\ref{A1bis}) are 
listed in section \ref{lin} and in appendix \ref{W3APP}.
\begin{figure}
\bc
\begin{picture}(0,0)%
\includegraphics{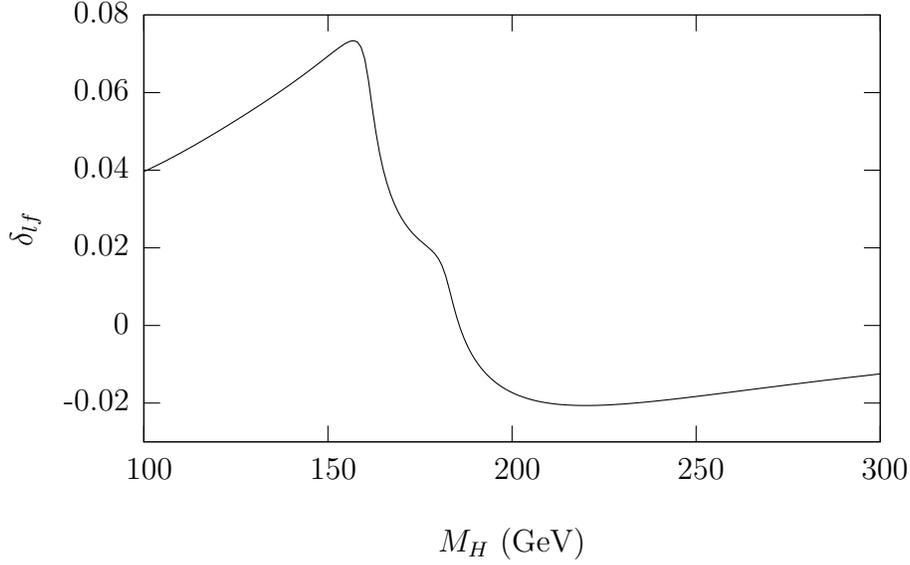}%
\end{picture}%
\begingroup
\setlength{\unitlength}{0.0200bp}%
\begin{picture}(18000,10800)(0,0)%
\put(3025,2932){\makebox(0,0)[r]{\strut{}-0.02}}%
\put(3025,4395){\makebox(0,0)[r]{\strut{} 0}}%
\put(3025,5859){\makebox(0,0)[r]{\strut{} 0.02}}%
\put(3025,7323){\makebox(0,0)[r]{\strut{} 0.04}}%
\put(3025,8786){\makebox(0,0)[r]{\strut{} 0.06}}%
\put(3025,10250){\makebox(0,0)[r]{\strut{} 0.08}}%
\put(3300,1650){\makebox(0,0){\strut{} 100}}%
\put(6769,1650){\makebox(0,0){\strut{} 150}}%
\put(10238,1650){\makebox(0,0){\strut{} 200}}%
\put(13706,1650){\makebox(0,0){\strut{} 250}}%
\put(17175,1650){\makebox(0,0){\strut{} 300}}%
\put(1050,6225){\rotatebox{90}{\makebox(0,0){\strut{} $\delta_{lf}$ }}}%
\put(10237,275){\makebox(0,0){\strut{} $M_{H}$ (GeV) }}%
\end{picture}%
\endgroup
\caption{\label{fig1} $\delta_{lf}$ as defined in Eq.~(\ref{delf}).}
\ec
\end{figure}

Writing
\be
\sigma(gg \to H) = \sigma_0 \, ( 1 + \delta_{lf} ) \, ,
\ee
where $\sigma_0$ is:
\be
\sigma_0 = \frac{G_F \alpha_S^2}{512 \, \sqrt{2} \, \pi} \, 
| {\mathcal G}^{1l} |^2 \, , \qquad 
{\mathcal G}^{1l} = {\mathcal G}_t^{1l} + {\mathcal G}_b^{1l} \, .
\ee
we have for $\delta_{lf}$:
\be
\delta_{lf} = \frac{2 \alpha}{| {\mathcal G}^{1l} |^2} 
             \left[ {\mathcal Re} ( {\mathcal G}^{1l}) \, 
	            {\mathcal Re} ( {\mathcal G}^{2l}_{lf})
	          + {\mathcal Im} ( {\mathcal G}^{1l}) \, 
		    {\mathcal Im} ( {\mathcal G}^{2l}_{lf})
		  \right] 
\, .
\label{delf}
\ee
In Fig.~\ref{fig1} we plot $\delta_{lf}$ computed with the VW routines and some
numerical values are collected in Table~\ref{tab1}.  The set of parameters used is the
following:
\bea
& & m_t = 173.1 \, \mbox{GeV} \, , \quad 
    m_b = 4.6 \, \mbox{GeV} \, ,  \quad
    m_W = 80.398 \, \mbox{GeV} \, , \quad
    \Gamma_W = 2.141 \, \mbox{GeV} \, ,  \nn\\
& & 
    m_Z = 91.1876 \, \mbox{GeV} \, , \quad
    \Gamma_Z = 2.4952 \, \mbox{GeV} \, , \nn\\
& & \alpha = 1/128 \, , \quad G_F = 1.16637 \cdot 10^{-5} \, \mbox{GeV}^{-2} 
\, , \quad \sin^2 \theta_W = 0.23149
\eea

\begin{table}[t]
\begin{center}
\begin{tabular}{|c|c|||c|c|||c|c|||c|c|||c|c|}
\hline
$m_H$  & $\delta_{lf}$ & $m_H$  & $\delta_{lf}$ 
& $m_H$  & $\delta_{lf}$ & $m_H$ & $\delta_{lf}$ & $m_H$ & $\delta_{lf}$ \\
\hline
110 &  0.04445 & 180 & 0.01723  & 250 & -0.01830  & 320 & -0.01047  & 390 & -0.00462 \\
\hline
120 &  0.04992 & 190 & -0.00888  & 260 & -0.01711  & 330 & -0.00952   & 400 & -0.00419 \\
\hline
130 & 0.05585  & 200 & -0.01729  & 270 & -0.01590  & 340 & -0.00860  & 410 & -0.00382 \\
\hline
140 & 0.06227  & 210 & -0.02003  & 280 & -0.01472  & 350 & -0.00754  & 420 & -0.00350 \\
\hline
150 & 0.06939  & 220 & -0.02063  & 290 & -0.01358  & 360 & -0.00655  & 430 & -0.00322 \\
\hline
160 & 0.06862  & 230 & -0.02025  & 300 & -0.01249  & 370 & -0.00576  & 440 & -0.00297 \\
\hline
170 & 0.02764  & 240 & -0.01939  & 310 & -0.01145  & 380 & -0.00514  & 450 & -0.00275 \\
\hline
\end{tabular}
\caption{$\delta_{lf}$ as a function of the Higgs boson mass ($m_H$ in
GeV). \label{tab1}}
\end{center}
\end{table}

\section{Conclusions}

In this paper we analyzed the set of GHPLs of a single variable containing
square roots in the weights. After recalling the definition and basic 
properties of the HPLs, we introduced the GHPLs with weights belonging to the
set $\{ -1-r,-r,-4,-1,0,1,4,r,1+r,c,\bar{c} \}$. This specific set of GHPLs
appears in the analytic expressions of the MIs that enter into the calculation
of the electroweak form factor \cite{Aglietti:2003yc,Aglietti:2004tq}. 

One of the main observations of the paper lies in the fact that, once the
weights are allowed to be complex,  the GHPLs with square roots in the weights
can be ``linearized'', {\it i.e.} expressed as a combination of GHPLs with
linear weights. These linearized GHPLs are functions of a transformed variable,
that is not unique, but can be properly chosen depending on the nature of the
weights.  The set $\{ -1-r,-r,-4,-1,0,1,4,r,1+r,c,\bar{c} \}$ can be linearized,
once and for all, with just two variable transformations.

The other observation concerns the possibility of a fast and precise numerical
evaluation of the linearized GHPLs using already existing numerical routines. In
particular, the {\tt C++/GiNaC} routines by Vollinga and Weinzierl 
\cite{Weinzierl} offer a well suited tool for this goal. 

Finally, the strategy for the numerical evaluation of GHPL presented in the 
paper is applied to the known case of electroweak light-fermion NLO corrections 
to the Higgs production in gluon fusion. 
We evaluate the GHPLs with square roots using the VW numerical routines. As a
further refinement,  while in \cite{Aglietti:2004nj} the corrections were
evaluated neglecting the effects of the $W$ and $Z$ widths, we consider here the
case  of complex $m_W$ and $m_Z$, getting a more realistic result.
It is worth to notice that the GHPLs with square roots allow for a very compact 
analytic expression of the results, which would be extremely lengthy if
expressed in terms of the linearized GHPLs.

\subsection*{Acknowledgments}

The algebraic manipulations of the paper were done using FORM \cite{FORM}.
R.~B. would like to thank C.~Studerus for useful discussions. The work of
R.~B. is supported by the Theory-LHC-France initiative of CNRS/IN2P3.
The work of G.~D. and A.~V. was supported by the
European Community's Marie-Curie Research Training Network under contract
MRTN-CT-2006-035505 (HEPTOOLS).

\appendix

\section{Some Examples at Weight 2 \label{W2APP}}

As simple examples, in this appendix we apply the procedure outlined in the 
paper to the following GHPLs at weight 2: $G(-r,-r;x)$ and $G(-r,-1;x)$. 
We find:
\bea
G(-r,-r;x) & = & \int_0^x \frac{dt}{\sqrt{t(t+4)}} G(-r;t) =
- \int_1^{\xi} \frac{d \eta }{\eta} \, \left[ - G(0;\eta) \right] 
\, , \nn\\
& = & \frac{1}{2} G(0;\xi)^2 \, , 
\label{Hmrmr} \\
G(-r,-1;x) & = & \int_0^x \frac{dt}{\sqrt{t(t+4)}} G(-1;t) =
- \int_1^{\xi} \frac{d \eta }{\eta} \, \left[
G(c;\eta) + G(\bar{c};\eta) - G(0;\eta) \right] \, , \nn\\
& = & - \frac{1}{3} \zeta(2) - G(0,c;\xi) - G(0,\bar{c};\xi) 
+ \frac{1}{2} G(0;\xi)^2 
\, .
\eea

\section{Some Examples at Weight 3 \label{W3APP}}

In this appendix, we provide the expressions for the 5 GHPLs at weight 3 
containing square roots in the weights, involved in the corrections of 
\cite{Aglietti:2004nj}. We have:
\bea
G(-r,-r,-r;x) & = & \int_0^x \frac{dt}{\sqrt{t(t+4)}} G(-r,-r;t) =
- \int_1^{\xi} \frac{d \eta }{\eta} \, \frac{1}{2} G(0;\eta)^2 \, , \nn\\
& = & - \frac{1}{6} G(0;\xi)^3 \, , \\
G(0,-r,-r;x) & = & \int_0^x \frac{dt}{t} G(-r,-r;t) =
\int_1^{\xi} d \eta \, \left( - \frac{1}{\eta} + \frac{2}{\eta-1} \right) 
\, \frac{1}{2} G(0;\eta)^2 \, , \nn\\
& = & 2 \zeta(3)
          - \frac{1}{6} G(0;\xi)^3
          + G(0;\xi)^2 G(1;\xi)
          + 2 G(0,0,1;\xi) \nn\\
& &   
          - 2 G(0;\xi) G(0,1;\xi)
\, , \\
G(-r,0,-1;x) & = &   \int_0^x \frac{dt}{\sqrt{t(t+4)}} G(0,-1;t) \, , \nn\\
& = & - \int_1^{\xi} \frac{d \eta }{\eta} \, \biggl[
            \zeta(2)
          - G(0,\bar{c};\eta)
          - G(0,c;\eta)
          + \frac{1}{2} G(0;\eta)^2 \nn\\
& &   
          - 2 G(1;\eta) G(0;\eta) 
	  + 2 G(0,1;\eta)
          + 2 G(1,\bar{c};\eta)
          + 2 G(1,c;\eta)
\biggr] \, , \nn\\
& = & 
       - \frac{10}{3} \zeta(3)
          + 2 K_1
          - \zeta(2) G(0;\xi)
          - \frac{1}{6} G(0;\xi)^3
          + G(0,0,\bar{c};\xi) \nn\\
& &  
          + G(0,0,c;\xi)
          - 2 G(0,1,\bar{c};\xi)
          - 2 G(0,1,c;\xi) \nn\\
& &  
          + 2 G(0;\xi) G(0,1;\xi)
          - 2 G(0,0,1;\xi)
\, , \\
G(-r,-r,-1;x) & = &    \int_0^x \frac{dt}{\sqrt{t(t+4)}} G(-r,-1;t) \, , \nn\\
& = & \int_1^{\xi} \frac{d \eta }{\eta} \, \Biggl[
\frac{1}{3} \zeta(2) + G(0,c;\eta) + G(0,\bar{c};\eta) - G(0,0;\eta) \Biggr] \, , \nn\\
& = & \frac{2}{3} \zeta(3)
          + \frac{1}{3} \zeta(2) G(0;\xi)
          - \frac{1}{6} G(0;\xi)^3
          + G(0,0,c;\xi)\nn\\
& &  
          + G(0,0,\bar{c};\xi)
\, , \\
G(-4,-r,-1;x) & = &  \int_0^x \frac{dt}{t+4} G(-r,-1;t) \, , \nn\\
& = & \int_1^{\xi} d \eta \, \left( \frac{1}{\eta} - \frac{2}{\eta+1} \right) 
\! \Biggl[
\frac{1}{3} \zeta(2) + G(0,c;\eta) + G(0,\bar{c};\eta) 
- G(0,0;\eta) \Biggr] \, , \nn\\
& = & 
       - \frac{5}{6} \zeta(3)
          - 2 K_2
          - \frac{2}{3} \zeta(2) G(-1;\xi)
          - 2 G(-1;\xi) G(0,\bar{c};\xi) \nn\\
& &
          - 2 G(-1;\xi) G(0,c;\xi)
          + G(-1;\xi) G(0;\xi)^2
          + 2 G(0,\bar{c},-1;\xi) \nn\\
& &
          + 2 G(0,c,-1;\xi)
          + \frac{1}{3} \zeta(2) G(0;\xi)
          - \frac{1}{6} G(0;\xi)^3 
          + 2 G(0,-1,\bar{c};\xi) \nn\\
& &
          + 2 G(0,-1,c;\xi)
          - 2 G(0;\xi) G(0,-1;\xi)
          + G(0,0,\bar{c};\xi) \nn\\
& &
          + G(0,0,c;\xi)
          + 2 G(0,0,-1;\xi)
\, .
\eea

In the formulas above, we introduced the two constants $K_1$ and $K_2$. They 
have a cumbersome expression in terms of known transcendental constants, that 
we omit here. Their numerical value is known with infinite precision and it is:
\bea
K_1 & = & \hspace*{3.3mm} 0.278425076639727748441973590814.. \, , \\
K_2 & = & - 0.152226248227607546589100778278.. \, .
\eea


\end{document}